# Stirling Stuff

*Dr John S. Reid, Department of Physics, Meston Building, University of Aberdeen, Aberdeen AB12 3UE, Scotland*

## Abstract

Robert Stirling's patent for what was essentially a new type of engine to create work from heat was submitted in 1816. Its reception was underwhelming and although the idea was sporadically developed, it was eclipsed by the steam engine and, later, the internal combustion engine. Today, though, the environmentally favourable credentials of the Stirling engine principles are driving a resurgence of interest, with modern designs using modern materials. These themes are woven through a historically based narrative that introduces Robert Stirling and his background, a description of his patent and the principles behind his engine, and discusses the now popular model Stirling engines readily available. These topical models, or alternatives made 'in house', form a good platform for investigating some of the thermodynamics governing the performance of engines in general.

-------------------------------------------------------------------------------------------------------------

## 1. Introduction

2016 marks the bicentenary of the submission of Robert Stirling's patent that described heat exchangers and the technology of the Stirling engine. James Watt was still alive in 1816 and his steam engine was gaining a foothold in mines, in mills, in a few goods railways and even in pioneering 'steamers'. Who needed another new engine from another Scot? The Stirling engine is a markedly different machine from either the earlier steam engine or the later internal combustion engine. For reasons to be explained, after a comparatively obscure two centuries the Stirling engine is attracting new interest, for it has environmentally friendly credentials for an engine. This tribute introduces the man, his patent, the engine and how it is realised in example models readily available on the internet.

## 2. Robert Stirling

Robert Stirling was a farmer's son, born in 1790 in rural Perthshire, the third in what would become a family of 8 children that included 5 girls. It was an age when agricultural improvements by way of land reclamation and new machinery were in full swing in Britain. His grandfather Michael Stirling had invented an early rotary threshing machine driven by water power; his younger brother James Stirling (born in 1799) would make his name as an engineer and it is a fair bet that family interests in the new machinery of the age inspired the young Robert into becoming an amateur mechanic. It was said that the girls in the family were also mechanically talented[1]. The early 19th century was also an age when a farmer's son attending one of the East of Scotland's four universities was a common-place occurrence. Authors disagree on the extent of Robert Stirling's formal education but records show that he enrolled in Edinburgh University in 1805 and took their courses for at least two years[2]. His second year included mathematics under the aegis of John Leslie, well-known for his study of heat[3]. It is not clear if he was later taught Natural Philosophy by the notable and enthusiastic John Playfair who would have given Stirling a wide-ranging contemporary understanding of the principles behind the operation of many kinds of machine[4].

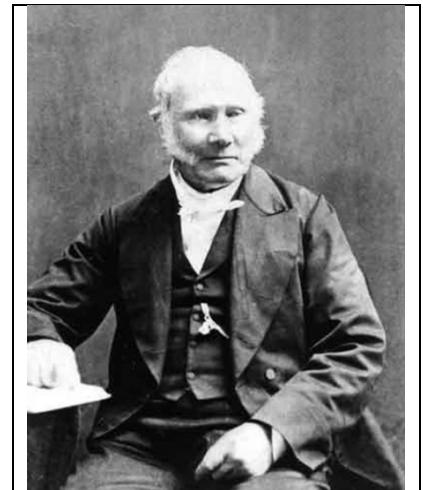

Robert Stirling (Fig. 1) would eventually have a long career as a Minister of the Church of Scotland, a graduate profession even two centuries ago. He spent additional time in divinity studies, perhaps at both the Universities of Edinburgh and Glasgow. Although he didn't formally collect an AM degree he was ordained in 1816 as Minister for Kilmarnock. He was known to maintain a workshop and a continuing interest in machinery. Indeed, even later in his career "*he sometimes surprised his neighbours at midnight by the hammering resounding from the anvil in his little smithy adjoining the manse*[5]".

1816 is noted even now as the disastrous '*year without a summer*' following the enormous eruption of Mount Tambora the previous year. Maybe the bad weather encouraged Stirling to spend even more time over the furnace in his smithy. Towards the end of 1816 he applied for his now famous patent[6] (No. 4081) for an 'Economiser' that discussed various ways of exchanging heat. This was well before the time that heat was recognised as a transfer of internal energy between bodies. Heat was then considered as a substance in its own right – 'caloric' that could flow between bodies or simply evaporate as radiation.

*Fig. 1 Robert Stirling in his later years.*

## 3. The patent of 1816

Stirling's 1816 patent began by describing a heat exchanger[7]. He suggested that one version could consist of two parallel channels in contact. Hot fluid passed down one channel and cooled as it did so; cool fluid passed in the opposite direction and heated up as it did so (Fig. 2).

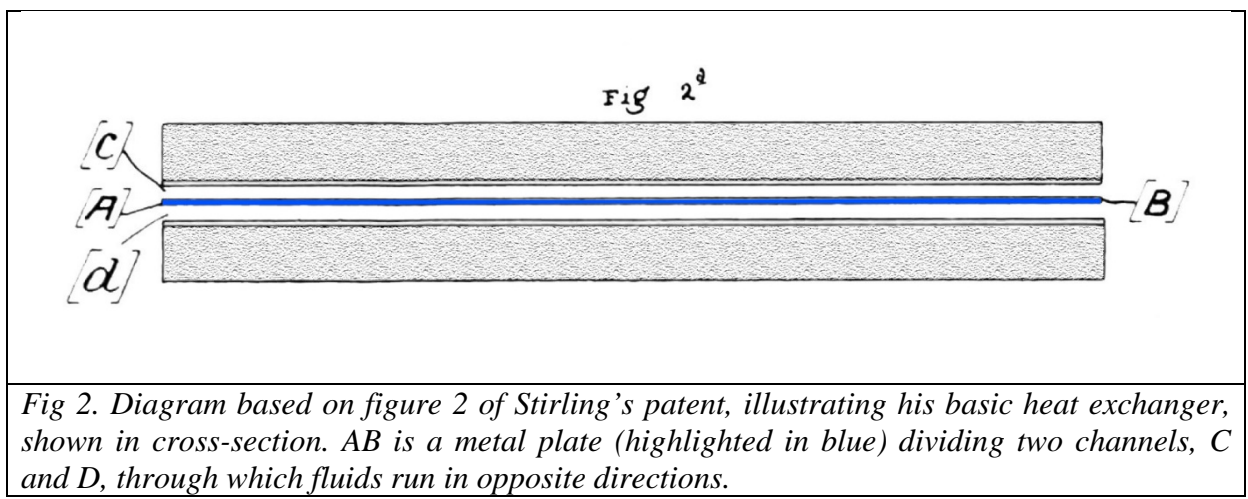

*Fig 2. Diagram based on figure 2 of Stirling's patent, illustrating his basic heat exchanger, shown in cross-section. AB is a metal plate (highlighted in blue) dividing two channels, C and D, through which fluids run in opposite directions.*

This is just the arrangement that should be built into all houses in cool climates for exchanging the internal heat of the stale air in a house with the cool ventilating air from outside. Sadly, 200 years after Stirling, very few houses include this heat exchange system. It's quite surprising that the simple heat exchanger hadn't been patented before then. The next version described a furnace with two flues. Air is drawn in through one flue and the combustion gases expelled through the other, which of course heats up. The roles of the flues are then reversed so the air intake is now through the heated flue and the hot gases will heat up the other flue. This makes the furnace more efficient since it is then fed with hot air and some of the exhaust gas heat is used - in modern parlance, an exhaust heat recovery system. This got Stirling thinking about heating and cooling gases and he went on to describe his engine.

We are all familiar now with the internal combustion engine using petrol, diesel, gas or some inflammable fuel. Such engines are a product of the second half of the $19^{th}$ century. When the hot gas has done its work in the piston it is expelled as exhaust and the piston can move relatively unimpeded. The steam engine is an external combustion engine in which the working fluid, the steam, passes through the engine and is likewise exhausted. The Stirling engine was a radically new invention, different from either of these. It is an engine where the working gas is sealed in the piston and operates between a hot 'reservoir' and a cold reservoir. The hot end can be maintained by combustion but nowadays it can also be maintained by other means such as solar heating. Indeed, a single engine can be designed readily that is not limited to one particular fuel. The Stirling engine is also attracting a lot of attention these days because the external combustion can be optimised to produce minimum waste products like carbon monoxide and unwanted nitrous oxides: basically, cleaner emissions that minimise its environmental footprint.

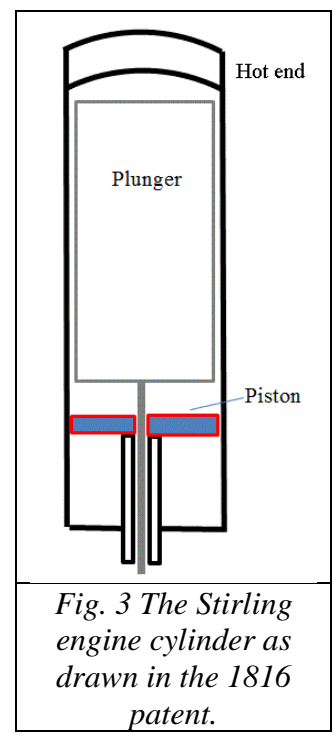

*Fig. 3 The Stirling engine cylinder as drawn in the 1816 patent.*

So, how does the engine work? The 1816 version shows why the original idea was a development of heat exchanging. Fig. 3 is re-drawn from part of Stirling's figure that shows a complete stationary engine, driving a shaft and flywheel 'for moving machinery'. The cylinder contains two independently moving parts: the tightly fitting power piston and a loosely fitting 'plunger'. The top of the cylinder is hot while the foot is cooled by air or water. The plunger is the heat exchanger. In Stirling's original concept the plunger was made of iron plate so it would quickly take in heat and give it out. When the bottom space is near its largest and coolest, the plunger comes down, moving air up its sides and reheating it due to the temperature gradient up its length. The air therefore arrives at the top needing less heat than it would do if just injected cold from outside. When the plunger rises it pushes hot air down its sides into the power space, in the process reheating the plunger. So during the cycle the plunger acts as a heat exchanger, without losing any energy to the outside. I'm using the word 'heat' here as a synonym for thermal energy transferred without any accompanying work.  The process has been likened to a heat sponge that alternately absorbs and releases thermal energy.  This action of Stirling's plunger he called 'regeneration'.  A feature of Stirling's original design is that the power piston is separated from the furnace heat by the plunger.  It is fed by hot gas but does not need piston rings that have to withstand cylinder walls at furnace temperature.

The Stirling engine was improved from its 1816 incarnation by Robert and his brother James, who jointly submitted two subsequent patents.  The 1827 patent (no. 5456) promoted a major development in that the power piston and displacer were now designed to work in separate cylinders.  This was the beginning of a wide range of variants that recognised that Stirling's original plunger did two jobs.  It displaced air, or whatever gas is enclosed, between the two ends and it acted as a heat exchanger.  These two jobs can be done separately by a 'displacer' and a 'regenerator' that may well be different components.  Taken to extremes, the hot and cold ends can be in separate chambers connected by a tube.  The displacer moves air between the tubes and the regenerator, if present in the tube, is the heat exchanger.  Whatever the configuration, by moving the air within the engine the displacer maintains the cyclic motion of the engine, which isn't created simply by having steady hot and cold ends to a cylinder.  In other engines the cycle is kept going by the engine's valves.  Like valves, the displacer is generally driven by the power train.  Stirling achieved this by connecting the plunger to the engine flywheel via levers.

The Stirling engine is not primarily about shunting heat around, to use the vernacular, but about producing work from the difference in temperature between hot and cold 'sources'.  The modern view is that internal energy is disorganised energy, work done represents organised energy.  Heat engines convert internal energy in hot sources into work but not 100% efficiently.  The underlying theory of engine efficiency was developed by Sadi Carnot a decade or so after Stirling's first patent.  Carnot recognised that an engine always heats up its coolant, thereby 'wasting' some of the initial heat energy.  Carnot deduced that an engine's efficiency necessarily depends on the temperature difference between its hot gas and its cold gas.  The bigger the difference, the more efficient an engine is in creating work from energy taken in as heat.  This is equally true of the Stirling engine.

## 4. The indicator diagram

At the end of the 18$^{th}$ century James Watt had developed the *indicator diagram* to determine the amount of work done by an engine.  This diagram is a plot of the cylinder pressure against the volume of the working substance.  Watt devised a method of making the engine produce its own diagram automatically.  A pressure gauge connected to the cylinder controlled the vertical movement of a pencil marking a sheet of paper and the cyclic motion of the piston, which is proportional to the volume, controlled the horizontal movement of the pencil.  As the engine went through each cycle the pencil traced out a loop and the work supplied was proportional to the area within the loop.  It was ingenious, and based on sound physics.

Fig. 4 shows the indicator diagram commonly quoted for an ideal Stirling engine. This diagram is a figment of textbooks. The perceptive J. Macquorn Rankine analysed the theory of thermodynamic engines in an extensive paper of 1854[8] in which he discussed the regenerator but he did not give a diagram for the Stirling engine. It appears that it was over 50 years after Stirling described his engine before the first theoretical analysis was given by Gustav Schmidt of the German Polytechnic Institute of Prague in 1871. Schmidt took the ideal cycle shown in Fig. 4, assumed sinusoidal variations of the volumes and an adjustable phase difference between the power piston and the regenerator. Even this analysis is not trivial[9] but it doesn't represent accurately a real Stirling engine.

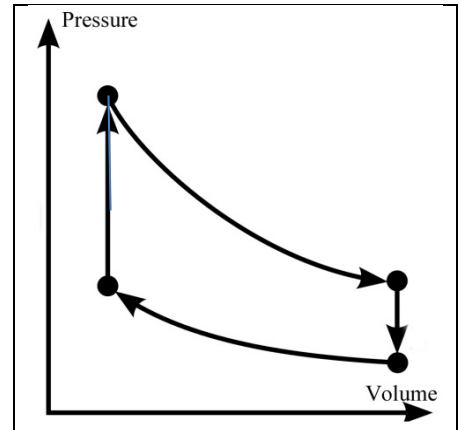

*Fig. 4 The indicator diagram for an ideal Stirling engine.*

The ideal cycle of the diagram is 'reversible' and therefore as efficient as an engine can be. The maximum efficiency of an engine is achieved by a 'Carnot engine', as described by Sadi Carnot. Efficiency is this context is defined as the work provided by the engine divided by the heat extracted from the hot source. The ideal Stirling engine has the same efficiency as a Carnot engine working between the same temperature limits, though its changes in pressure and volume are different from the usually quoted 'Carnot Cycle'. The real Stirling engine involves forcing the gas within it rapidly through the narrow channels of a heat exchanger. This process involves friction and is never going to be reversible. Another problem with the Stirling engine is that all the gas inside isn't at the same temperature and pressure, making it more complex than a textbook idealisation. In real life Stirling engines usually can't manage half the efficiency of the ideal engine[10]. Robert Stirling didn't design his engine from any theory (there was no useful theory of engines in 1816): just an inspired idea supported by his own mechanical experience. Of course theory can be useful but the over-simplification of the Schmidt model rather distracts from what is happening in the real insides of a Stirling engine.

All that said, the indicator diagram is at least a raft in a sea of practical difficulties. Starting at the top left and going round clockwise, the power stroke is provided by the expansion of hot gas, during which heat is also taken in, ideally at constant temperature. The bottom line represents the compression of the cold gas, during which some heat is given out to the coolant, again ideally at constant temperature. This compression uses up some but not all of the work generated by the expansion and in practice is driven by the flywheel. The net result is that the work supplied by the engine is represented by the area within the diagram, as usual. The vertical sections are the least realistic parts, describing motion of the gas past the regenerator, ideally at constant volume.

One feature suggested by the ideal indicator diagram is that if the cylinder is filled with a gas at high pressure instead of the ordinary atmospheric pressure that Stirling himself described initially then a change in pressure of say 60% during the cycle will result in a greater absolute change in pressure and hence more work will be done in each cycle. Modern Stirling engines favour the use of gas at high pressure. The Stirling engine may not live up to its 'theoretical' efficiency but with a high temperature provided by solar energy, the Stirling engine coupled to an alternator can still produce electricity with greater efficiency than a typical solar cell. This is one of the reasons Stirling engines are now being developed seriously.

With the indicator diagram still in view it is worth saying that when a Stirling cycle is driven in reverse (i.e. anticlockwise round the diagram) by an external motor, the engine acts as a heat pump or a refrigerator. Having no volatile refrigerant like ammonia or CFCs, it is environmentally clean and can operate at temperatures down to liquid nitrogen (about -200° C).

## 5. The patent of 1827

Robert and his brother James submitted a new patent in 1827 that described in considerable detail a substantial Stirling engine that had been built under James Stirling's guidance. This contained several important developments. First, they placed the displacer and power piston in separate cylinders, as mentioned. This has come to be known as the alpha ($\alpha$) configuration. The 1816 geometry with displacer and piston in the same cylinder is the beta ($\beta$) version. They took a leaf out of James Watt's book and made the engine 'double acting', meaning that it produced power on both the upstroke and downstroke. Watt did this by valves; the Stirling's by arranging two displacers, one to feed the foot of the power cylinder and one the top. They inverted the design so that the hot ends of the displacers were underneath the machinery and they added a compressed air pump so the air within could be increased in pressure to around 20 atmospheres.

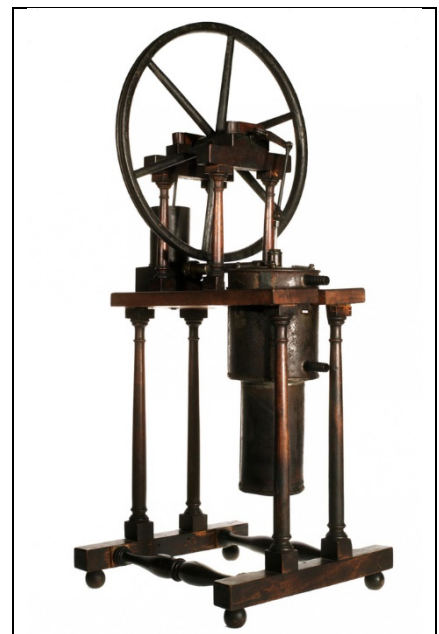

*Fig. 5 Stirling's own model engine, given to the University of Glasgow. Courtesy Hunterian Museum.*

All these were very intelligent developments and if their design had been as good in practice as on paper then the Stirling engine could have been a serious competitor to steam. It was more efficient and hence more economical, and safer, an important point in an age when steam boiler explosions and engine failures were common. James Stirling's engine generated about 40 horsepower, enough to turn many machines in the foundry in Dundee he managed. It was James Stirling himself who found the big practical problem. In the days before Bessemer's steel, the cast iron available for the air vessel could not take red heat for an extended period and burnt out. He was let down by the materials of the day. The Stirlings did submit a third patent in 1840 (no. 8652) with technical improvements but Bessemer steel was still in the future.

## 6. Stirling engine models

Robert Stirling himself made at least two models of his engine. He gave one to the University of Edinburgh in the mid-1820s[11] and one to the University of Glasgow in 1828 (Fig. 5). If James Watt had personally made models of his innovative steam engine, or Niklaus Otto of the first practical internal combustion engine then they would be considered 'national treasures'. Stirling's models are becoming that. In 2015 the Glasgow model received the Institution of Mechanical Engineers (IMechE) Engineering Heritage Award. William Thomson, later Lord Kelvin, demonstrated this model to his students for decades. It incorporates the improvement of water cooling. An entire book gives further details of both models, including plans for the model maker to replicate them[12]. The London Science Museum displays a sectioned copy of the Edinburgh model. Nowadays, small, modern models for sale abound on the internet.

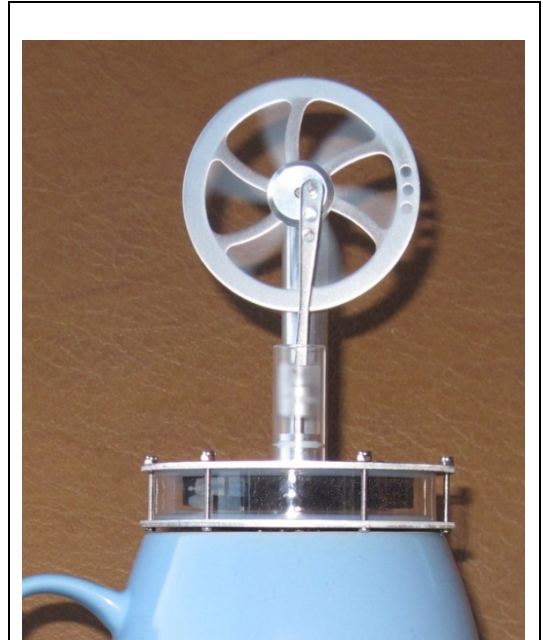

*Fig. 6 Small Stirling engine model of the γ-configuration sitting on a mug of hot water. The power piston is beneath the flywheel; the displacer is the dark object within the lower wide chamber.*

The Stirling engine scales down very well, making possible small models that really work, even for small temperature differences. In fact it is scaling up Stirling engines to produce powers greater than about 10 kW that is particularly difficult. The scaling problem arises because the power an engine could produce scales as the cube of its dimensions but the surface area through which externally generated heat can be conducted into the engine increases only as the square of the dimensions. The model shown in Fig. 6 is the so-called gamma (γ) version of the engine in which the displacer occupies a separate but directly connected chamber from the power piston. The displacer chamber is much wider than the power piston in order to transfer a reasonable amount of heat from the warm plate. The displacer stroke is correspondingly smaller.

Engines of this type are nicely described on the web, with animations[13]. The use of glass or Perspex for the cylinder sides helps to reduce the conduction of heat between the hot and cold reservoirs. Robert Stirling's use of a brass cylinder for the displacer in his first model explains why the Edinburgh version stopped after quite a short run. It was undoubtedly the reason why he added a cooler to the Glasgow model. The author's version never fails to impress, for it runs for well over an hour with its base, the hot reservoir, on a mug of hot water, tea or coffee. The top plate is the cool reservoir. It will run with the bottom plate warmed only by the heat of one's hand. Alternatively the model can be sat on a table and a cube of ice in a plastic bag (to contain the melted ice) placed on the top plate. There is no exhaust at all, no hiss of escaping steam nor throaty roar of fumes forced out into the atmosphere; just the quiet clickety-click of a few moving parts.

Models like this can make an excellent basis for project work, exploring how much work the engine can do and how efficient it is. Fig. 7 is an indicator diagram from a similar model to that shown in Fig. 6, determined by Hiroko Nakahara of the University of British Columbia from instrumenting his model. It shows very well that a real Stirling engine differs substantially from the 'ideal' engine. His figures show that their engine produced 3.4 mW of power with an 'ideal' efficiency of 19%. In Nakahara's model the displacer was operated by a crank run from the same spindle as the flywheel.

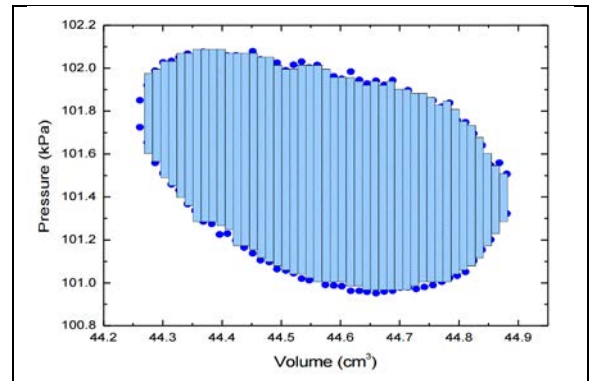

*Fig. 7 Indicator diagram measured for a model similar to that in figure 4: courtesy Hiroko Nakahara[14]*

In the author's model (from Starpower), the piston includes a very small magnet in its base and when it nears the bottom of its travel the magnet attracts the foam displacer. This quickly moves the cool air above it to the warm plate below. As the piston rises the displacer is quickly dropped, now moving the heated air up to the piston cylinder. The displacer is no longer moving sinusoidally and this should make for a slightly more efficient engine, though it will take instrumentation to verify this.

Fig. 8 shows a typical model of the alpha configuration of the Stirling engine in which the displacer is situated in a separate cylinder connected via piping that usually contains the regenerator. A wide variety of geometries can be found, from directly opposing cylinders, V-layout to the right-angle geometry of the model in Fig. 8. A small meths burner heats the displacer piston and the model runs quickly for as long as the flame lasts. The enterprising amateur mechanic can make an alpha configuration version from an existing single-cylinder internal combustion engine, but replacing the ignition with a coupled displacer cylinder highlights the added mechanical complexity of the Stirling engine. A variant of the α-version is the engine that James Stirling built for the foundry in Dundee he managed in the 1830s and in one form or another is the version that has been most commonly developed since for commercial engines that move machinery.

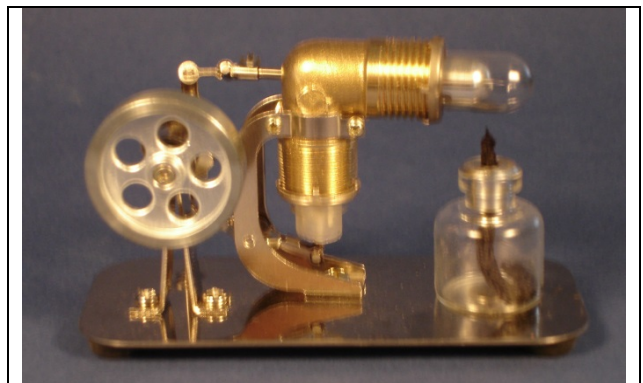

*Fig. 8 Stirling engine model of the α-configuration with separate but connected cylinders containing the power piston and displacer: courtesy Allan Mills.*

## 7. A modern take on the Stirling engine

The free-piston Stirling engine is a development credited to William Beale in the 1960s that takes us back to the original β-configuration. Beale founded Sunpower Systems, whose website has an animation of the basic idea[16]. Free piston engines have no rotary parts and can be completely sealed. A version ideal for a useful power generator is shown in Fig. 9. A nice piece of physics makes the operation possible. The phase angle between the displacer and the power piston can be controlled by adjusting the resonant frequency of the displacer and its attached spring. The piston is attached to a powerful magnet and the application of more basic physics results in its oscillatory motion inducing the generator's current in the enclosing coil. High pressure helium is sealed in. There is no need for lubrication, cleaning or other regular maintenance. The whole engine is a sophisticated compound of simple ideas. The challenge is to make such devices cheap enough to generate a few kilowatts of electrical power economically.

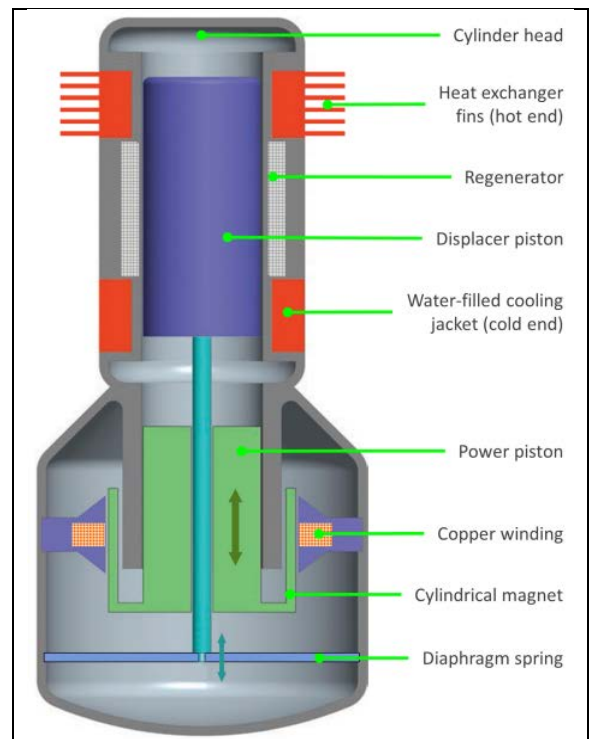

*Fig. 9 Sectional view of a free-piston Stirling engine as used in the Baxi Ecogen and Viessmann micro-CHP appliances: courtesy Energy Saving Advisor[15].*

Electricity can perform almost every function that needs power (provide transport, create heat and light, run machinery, etc.). It seems likely that the biggest role for Stirling engines in the future will be to create electricity for local use.

## 8. Robert Stirling and his legacy

Robert Stirling moved to the Parish of Galston in Ayrshire in 1824 where he was to serve as its well respected minister for 54 years. Retirement and a pension were not an option in the 19th century. This was close to Robert Burns country. His sons Patrick, William, Robert and James all became railway engineers: engineering was, it seems, in their genes. The fifth son, David, became a minister in Ayrshire. Five sons when he himself had had five sisters. Robert Stirling died at the age of 88 and is buried in Galston Parish churchyard where there is a memorial stone that was renewed in 2014 (Fig. 10). In the same year he was inducted into the Scottish Engineering Hall of Fame[17], joining outstanding engineers like James Watt, Thomas Telford, J W Macquorn Rankine and physicists Lord Kelvin and James Clerk Maxwell.

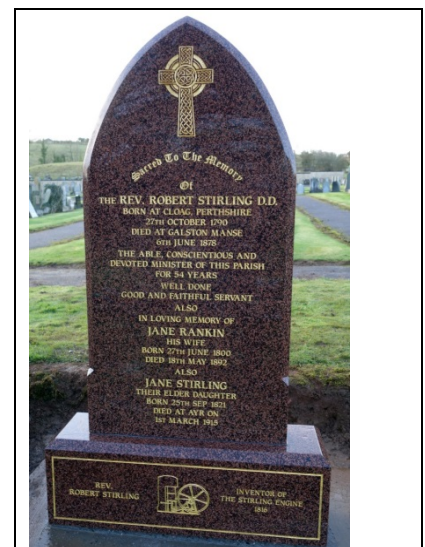

*Fig. 10 The new memorial stone in Galston cemetery: courtesy Galston Parish Church.*

The Stirling engine had several advantages going for it. These included potentially better efficiency than steam engines, no valves, no cylinder exhaust, quieter running, no need for a plentiful supply of water, no risk of scalding boiler explosions with the accompanying high insurance, operation using a wide range of fuel, and more. Ideal, you might have thought, for a locomotive. Yet the fact is that the Stirling engine made little impact in the 19th century. One of the main troubles was that the high temperatures desirable for good efficiency were beyond the metallurgy of the day. Another requirement for good power output is to use very high pressure for the gas within the engine. High temperatures also create trouble for lubrication and piston rings. Stirling engines were large and heavy for the horsepower they produced. In the early years, although Robert Stirling maintained a close interest in his idea it could not have helped that he had a full-time job in quite a different field. His brother James, engineer by trade and enthusiast, seemed to bow out of developments in the 1840s. One can speculate that it was James's connections that facilitated the University of St Andrews awarding Robert Stirling a doctorate in 1840[18].

It is also clear in hindsight that heat engines of the Stirling type needed development on an industrial scale to become effective. Such development went into steam engines but it was individuals who tried to develop the Stirling engine. To cite one example, John Ericsson was a Swede who spent much of his almost equally long life developing similar 'caloric' engines between the 1820s and 1880s, with displacers and regenerators. He was a confident man, ambitious, excellent at raising capital and fairly prolific with his designs. He worked for much of the time in England and the United States. His ideas spawned several companies and at least three statues to him exist. In spite of all this, his engines (which were a variant on Stirling's in that they had valves) failed to displace the less efficient, noisy, potentially dangerous steam engines.

There seems to be a feeling, though, that the 21st century is the time of the Stirling engine. Metallurgy and lubrication are no longer insoluble problems. We shall surely hear more about Stirling engines in the coming years as their environmentally favourable credentials receive greater prominence. The web has many links to designs for small Stirling engines created by a wide spectrum of the interested from tinkerers to NASA, some with You-Tube videos. You can also find computer programs for predicting thermodynamic behaviour and efficiency, the Stirling Engine Society and other forums for enthusiasts.

## 9. Acknowledgement

I would like to thank Allan Mills for his encouragement to write this article.

**Footnotes**

---

[1] This comment is made in the 1876 Institution of Civil Engineers obituary for James Stirling, published in Grace's Guide to British Industrial History.